\documentclass[preprint2]{aastex}

\usepackage{graphicx}
\DeclareGraphicsExtensions{.ps}
\shorttitle{\indent \def Jets from Light Bridges} \shortauthors{Tian et al.}

\begin{document}

\title{Frequently Occurring Reconnection Jets from Sunspot Light Bridges}

\author{Hui Tian\altaffilmark{1}, Vasyl Yurchyshyn\altaffilmark{2,3}, Hardi Peter\altaffilmark{4}, Sami K. Solanki\altaffilmark{4,7}, Peter R. Young\altaffilmark{5,8,9},Lei Ni\altaffilmark{6},Wenda Cao\altaffilmark{2},Kaifan Ji\altaffilmark{6},Yingjie Zhu\altaffilmark{1}, Jingwen Zhang\altaffilmark{1}, Tanmoy Samanta\altaffilmark{1}, Yongliang Song\altaffilmark{1}, Jiansen He\altaffilmark{1}, Linghua Wang\altaffilmark{1}, Yajie Chen\altaffilmark{1}}
\altaffiltext{1}{School of Earth and Space Sciences, Peking University, Beijing 100871, China; huitian@pku.edu.cn}
\altaffiltext{2}{Big Bear Solar Observatory, New Jersey Institute of Technology, 40386 North Shore Lane, Big Bear City, CA 92314-9672, USA.}
\altaffiltext{3}{Korea Astronomy and Space Science Institute, 776 Daedeok-daero, Yuseong-gu, Daejeon, 305-348, Republic of Korea.}
\altaffiltext{4}{Max Planck Institute for Solar System Research, Justus-von-Liebig-Weg 3, 37077 G\"ottingen, Germany.}
\altaffiltext{5}{College of Science, George Mason University, Fairfax, VA 22030, USA.}
\altaffiltext{6}{Yunnan Observatories, Chinese Academy of Sciences, Kunming 650011, China.}
\altaffiltext{7}{School of Space Research, Kyung Hee University, Yongin, Gyeonggi-Do,446-701, Republic of Korea.}
\altaffiltext{8}{Code 671, NASA Goddard Space Flight Center, Greenbelt, MD 20771, USA.}
\altaffiltext{9}{Northumbria University, Newcastle Upon Tyne, NE1 8ST, UK.}

\begin{abstract}
Solid evidence of magnetic reconnection is rarely reported within sunspots, the darkest regions with the strongest magnetic fields and lowest temperatures in the solar atmosphere. Using the world's largest solar telescope, the 1.6-meter Goode Solar Telescope, we detect prevalent reconnection through frequently occurring fine-scale jets in the H${\alpha}$ line wings at light bridges, the bright lanes that may divide the dark sunspot core into multiple parts. Many jets have an inverted Y-shape, shown by models to be typical of reconnection in a unipolar field environment. Simultaneous spectral imaging data from the Interface Region Imaging Spectrograph show that the reconnection drives bidirectional flows up to 200~km~s$^{-1}$, and that the weakly ionized plasma is heated by at least an order of magnitude up to $\sim$80,000 K. Such highly dynamic reconnection jets and efficient heating should be properly accounted for in future modeling efforts of sunspots. Our observations also reveal that the surge-like activity previously reported above light bridges in some chromospheric passbands such as the H${\alpha}$ core has two components: the ever-present short surges likely to be related to the upward leakage of magnetoacoustic waves from the photosphere, and the occasionally occurring long and fast surges that are obviously caused by the intermittent reconnection jets.

\end{abstract}

\section{Introduction}

Magnetic reconnection, a physical process in which the magnetic field topology is rearranged and a part of the magnetic energy is converted into thermal and kinetic energy, is believed to be one of the most important energy release mechanisms in astrophysical, solar and space plasmas \citep[e.g.,][]{Priest2000,Deng2004,Gosling2005,Phan2006,Zhang2012,Wyper2017}. In the light of the proposal that the solar atmosphere is powered by prevalent small-scale magnetic reconnection \citep{Parker1988}, efforts have been made to search for evidence of small-scale reconnection events in the solar atmosphere in the past $\sim$30 years. Collimated jet-like features may be caused by upward propagating reconnection outflows, thus are often cited as evidence of reconnection \citep[e.g.,][]{Katsukawa2007,DePontieu2007,Tian2014a}. However, it has been suggested that these unidirectional collimated jets may also result from magnetohydrodynamic turbulence \citep{Cranmer2015}, warps in two-dimensional sheet-like structures \citep{Judge2011}, or amplified magnetic tension caused by ion-neutral interactions \citep{Martinez-Sykora2017}.

Meanwhile, high-resolution observations have revealed the frequent occurrence of jets with an inverted Y-shape in coronal holes \citep[e.g.,][]{Cirtain2007} and active regions outside sunspots \citep[e.g.,][]{Shibata2007,Singh2012,Yang2011,Shen2012,Tian2012,Zhang2014,Chen2015}. The inverted Y-shape appears to be a natural consequence of reconnection between small-scale magnetic bipoles (or a magnetic arcade) and the unipolar background fields \citep[e.g.,][]{Moreno-Insertis2013,Sterling2015}, thus providing strong evidence for magnetic reconnection. Such inverted Y-shaped structures have been rarely observed within sunspots, the darkest regions with the strongest magnetic fields and lowest temperatures in the solar atmosphere. Previously signatures of such structures in sunspots were identified only in a penumbral region \citep{Zeng2016} and a penumbral intrusion into the umbra \citep{Bharti2017}.

Some sunspots have light bridges, which are bright lanes dividing the dark umbra into multiple parts. Light bridges appear to comprise multi-thermal and multi-disciplinary structures extending beyond the photosphere \citep{Rezaei2017}. Previous chromospheric observations in the H${\alpha}$ and Ca~{\sc{ii}} passbands sometimes reveal long-lasting recurring surge-like (or jet-like) activity above light bridges, which is often described by different authors as H${\alpha}$ surges, plasma ejections, chromospheric jets, or light walls \citep[e.g.,][]{Roy1973,Asai2001,Shimizu2009,Bharti2007,Bharti2015,Yuan2016,Yang2016,Hou2016a,Hou2016b,Song2017}. These surges are usually suggested to be driven by magnetic reconnection, mostly based on their high speeds ($\sim$100 km~s$^{-1}$) and coincidence with strong currents in some observations \citep{Louis2014,Toriumi2015b,Robustini2016}. However, other authors reported low speeds ($\sim$15 km~s$^{-1}$) of light bridge surges and nearly stationary oscillating periods of a few minutes, suggesting their cause to be leakage of p-modes from the photosphere rather than reconnection \citep{Yang2015,Zhang2017}.

Using joint observations from the Goode Solar Telescope \citep[GST, previously called New Solar Telescope,][]{Cao2010}, the Interface Region Imaging Spectrograph \citep[IRIS,][]{DePontieu2014} mission and the Atmospheric Imaging Assembly \citep[AIA,][]{Lemen2012} instrument on board the Solar Dynamics Observatory \citep[SDO,][]{Pesnell2012}, we report detection of frequently occurring magnetic reconnection and the resultant significant heating in the lower atmosphere of sunspot light bridges. Our observations also solve the dispute on the nature of the persistent surge-like chromospheric activity observed above some light bridges.

\section{Observations and data reduction}

 \begin{table*}[]
\caption[]{Summary of the data analyzed in this paper.}\label{t1}
\begin{center}
\begin{tabular}{| p{1.8cm} | p{1.7cm} | p{1.8cm} | p{2.2cm} | p{1.8cm} | p{2.2cm} | p{1.5cm} | }
\hline Date  & Telescope & Time (UT) &  IRIS Pointing (x,y) & Passbands & Cadence in each passband & Pixel size \\
\hline 
 2014 Oct 29 & GST &  17:30--19:44 &  & H${\alpha}$ core & 30 s &  0.030$^{\prime\prime}$ \\
  &  &  &  & H${\alpha}$ --0.8 \AA{} & 30 s &  0.030$^{\prime\prime}$ \\
  &  &  &  & H${\alpha}$ +0.8 \AA{} & 30 s &  0.030$^{\prime\prime}$ \\
  &  &  &  & TiO & 30 s &  0.034$^{\prime\prime}$ \\
  \hline
 2014 Oct 29  & IRIS &  15:30--18:18 & (910$^{\prime\prime}$, --228$^{\prime\prime}$) & 2796 \AA{} & 16 s &  0.33$^{\prime\prime}$ \\
  &  &  &  & 1330 \AA{} & 16 s &  0.33$^{\prime\prime}$ \\
 \hline
 2014 Oct 28 & IRIS & 08:20--09:12 & (788$^{\prime\prime}$, --318$^{\prime\prime}$) & 2796 \AA{} & 39 s &  0.17$^{\prime\prime}$ \\
 &  &   &  & 1400 \AA{} & 39 s &  0.17$^{\prime\prime}$ \\
\hline
 2014 Oct 28 & SDO/AIA & 08:20--09:12 &  & 1700 \AA{} & 24 s &  0.6$^{\prime\prime}$ \\
 &  &   &  & 171 \AA{} & 12 s &  0.6$^{\prime\prime}$ \\
\hline
\end{tabular}
\end{center}
\end{table*}

We mainly analyze four datasets in this study: GST observation from 17:30 UT to 19:44 UT on 2014 Oct 29, IRIS observations from 15:30 UT to 18:18 UT on 2014 Oct 29, IRIS and AIA observations from 08:20 UT to 09:12 UT on 2014 Oct 28. Instrument pointing information, passband, cadence and pixel size of these datasets are summarized in Table~\ref{t1}. The observed sunspot was located in NOAA Active Region (AR) 12192, which was close to the west limb of the solar disk on these two days. This sunspot was also used by \cite{Yurchyshyn2017} for a white light flare study. The sunspot group in NOAA AR 12192 has a $\beta$$\gamma$$\delta$ type configuration. The analyzed sunspot was already in its mature stage from Oct 17 to Oct 30, when this flare-productive AR passed through the solar disk.

The 1.6-m GST in the Big Bear Solar Observatory (BBSO) has achieved nearly diffraction limited observations at a spatial resolution better than 0.1$^{\prime\prime}$, which are ideal for the study of small-scale physical processes in the photosphere and chromosphere. All scientific instruments in the Coud\'e Lab are currently in daily operation \citep{Cao2010}. The GST data were taken on 2014 Oct 29 in the passbands of TiO, H${\alpha}$ core, and H${\alpha}$ wings at --0.8  \AA{} and +0.8 \AA{}. The TiO 7057 \AA{} filter images were taken with the Broad-band Filter Imager (BFI), and they are mainly used to reveal underlying photospheric structures. The H${\alpha}$ data were taken with the Visible Imaging Spectrometer (VIS) of GST. VIS is based on a narrow-band tunable Fabry-P\'erot interferometer that offers a bandpass smaller than 0.08 \AA{}. The H${\alpha}$ core passband is a typical chromospheric passband, while the emission in the H${\alpha}$ wings at --0.8 \AA{} and +0.8 \AA{} appears to come from the upper photosphere or lower chromosphere. Horizontal and vertical noise patterns are present in the original H${\alpha}$ images. To remove this noise pattern, we have applied a two-dimensional discrete wavelet transform (DWT) and Principal Component Analysis (PCA) reconstruction technique. This procedure is described as follows:
(1) Decomposition of an image by 4th order Daubechies wavelet with 4 levels and generation of an approximation, horizontal, vertical and diagonal coefficients for each level.
(2) Application of the PCA reconstruction technique to the horizontal and vertical coefficients for each level to extract horizontal and vertical patterns of coefficients in the wavelet domain, respectively. 
(3) Recomposition of the extracted coefficients by an inverse DWT and generation of a horizontal and vertical noise pattern image in a spatial domain.
(4) Removal of the noise pattern image from the observed image.

The IRIS pointing was (910$^{\prime\prime}$, --228$^{\prime\prime}$) in the 2014 Oct 29 observation and (788$^{\prime\prime}$, --318$^{\prime\prime}$) in the 2014 Oct 28 observation, both very close to the limb. GST followed the IRIS pointing on 2014 Oct 29. Calibrated level 2 data of IRIS are used in our study. Dark current subtraction, flat field correction, and geometrical correction have all been taken into account in the level 2 data \citep{DePontieu2014}. The fiducial lines are used to achieve an alignment between images taken in different spectral windows and SJI filters. The 2796 \AA{} filter samples emission mainly from the Mg~{\sc{ii}} 2796 \AA{} spectral line that is formed in the upper chromosphere at a temperature of $\sim$10,000 K. The 1330 \AA{} and 1400 \AA{} filters sample emission of the strong C~{\sc{ii}} 1334 \AA{} and 1335 \AA{} spectral lines formed around 30,000 K and the Si~{\sc{iv}} 1394 \AA{} and 1403 \AA{} spectral lines formed around 80,000 K, respectively. Note that these temperatures refer to the formation temperatures of the lines under ionization equilibrium. In the case of non-equilibrium ionization, the transition region lines may sample significant emission from plasma with lower temperatures \citep{Olluri2015}.

We have also analyzed images taken in the 1700 \AA{}~and 171 \AA{}~passbands of SDO/AIA. The 1700 \AA{}~passband mainly samples the ultraviolet continuum emission formed around the temperature minimum region (TMR). The cadence and pixel size of the 1700 \AA{}~images are 24 s and $\sim$0.6$^{\prime\prime}$, respectively. The emission in the 171 \AA{}~passband mainly comes from the Fe~{\sc{ix}}~171.107 \AA{} line formed at a temperature of $\sim$80,0000 K and some other emission lines formed at typical transition region temperatures \citep{ODwyer2010}. The cadence and pixel size of the 171 \AA{}~images are 12 s and $\sim$0.6$^{\prime\prime}$, respectively. The coalignment between the AIA images and IRIS images are achieved by matching locations of some commonly observed small-scale dynamic events. 

\section{Reconnection jets and the associated heating: GST and IRIS observations on 2014 Oct 29}

\begin{figure*}
\centering {\includegraphics[width=\textwidth]{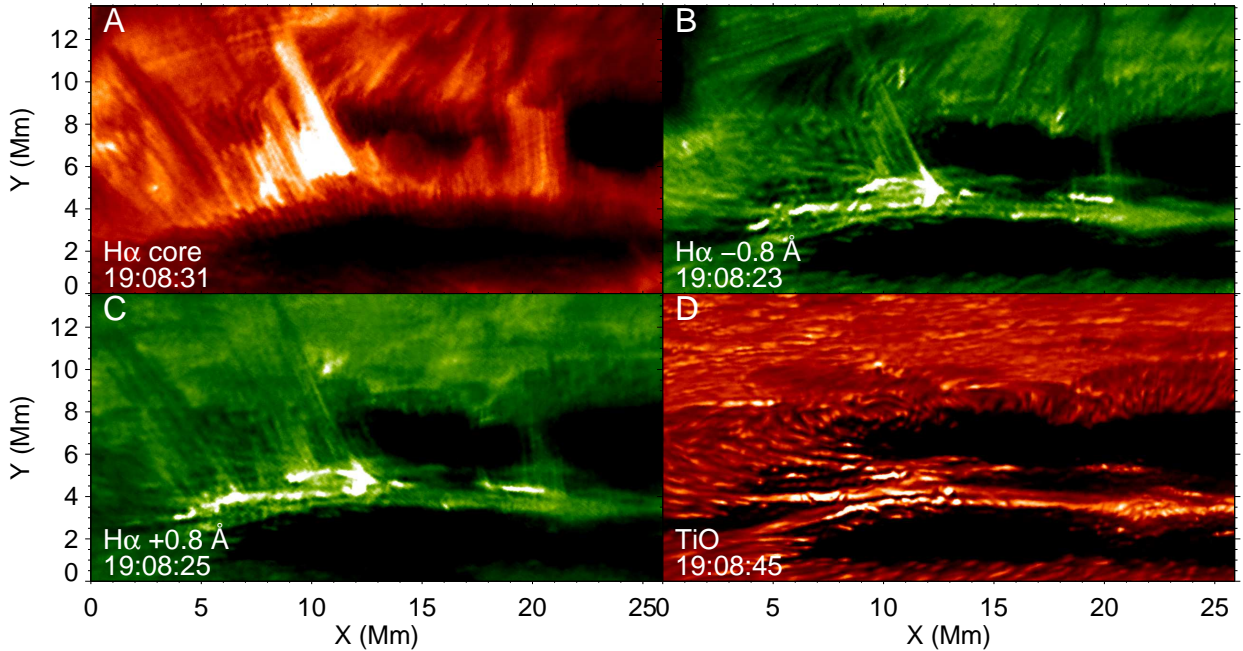}} \caption{ The sunspot light bridges observed by GST in the passbands of TiO, H${\alpha}$ core and H${\alpha}$ wings at $\pm$0.8 \AA{} at 19:08 UT on 2014 Oct 29. An associated animation (m1.mp4) is available online. } \label{fig.1}
\end{figure*}

\begin{figure*}
\centering {\includegraphics[width=\textwidth]{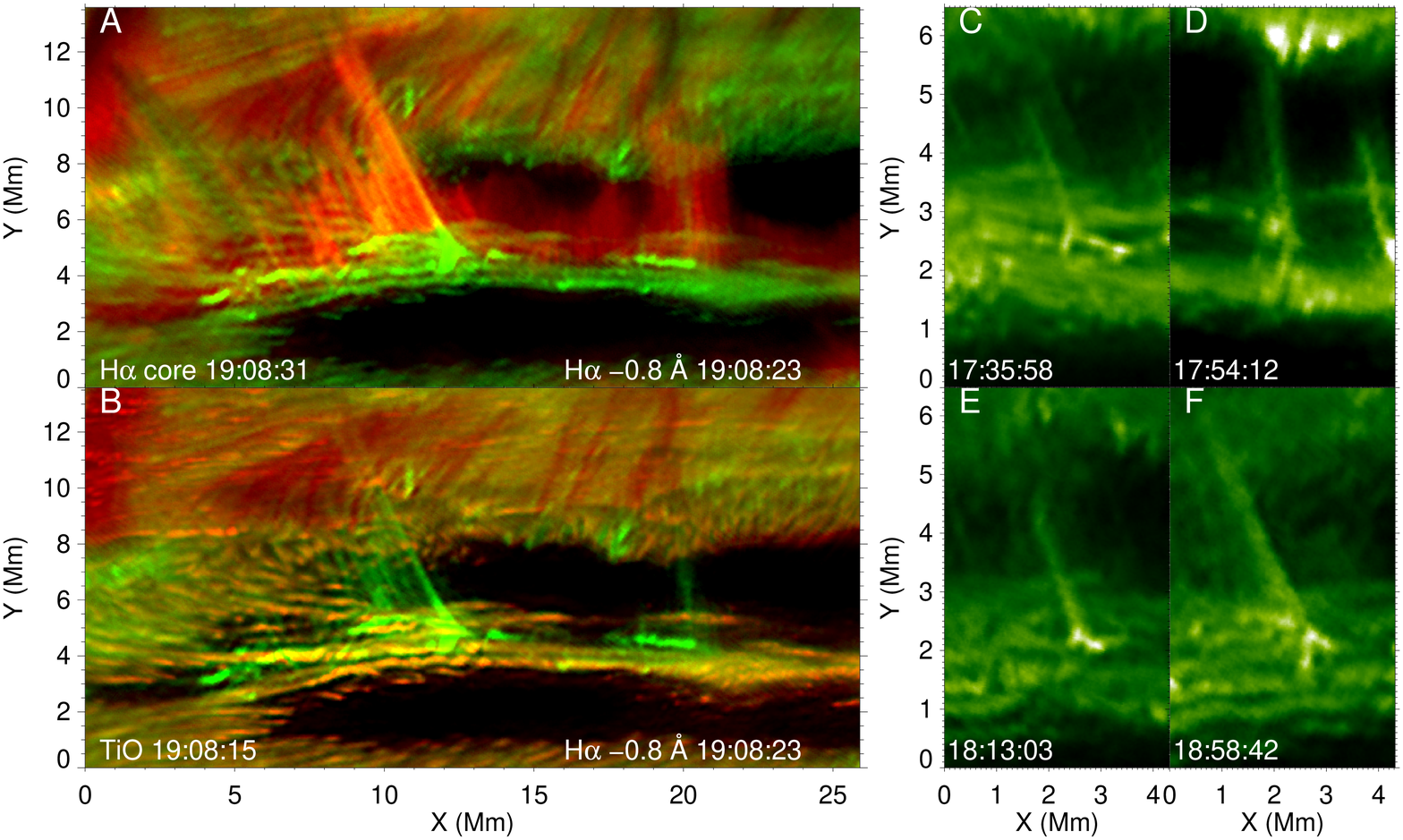}} \caption{ Examples of small-scale jets with an inverted Y-shape from the sunspot light bridges observed by GST on 2014 Oct 29. (A) Composite image of H${\alpha}$ core (red) and wing at --0.8 \AA{} (green) at 19:08 UT. (B) Composite image of TiO (red) and H${\alpha}$ wing at --0.8 \AA{} (green). (C-F) Images of H${\alpha}$ wing at --0.8 \AA{} taken at four different times. An associated animation (m2.mp4) is available online. } \label{fig.2}
\end{figure*}

Figure~\ref{fig.1} shows an example of images taken in the four passbands of GST mentioned above. There appear to be multiple closely lying filamentary light bridges within the sunspot umbra. The associated online animation reveals the prevalence of narrow and intermittent jets above the light bridges from the H${\alpha}$ off-band images at --0.8 \AA{} and +0.8 \AA{}. Significant brightenings are often observed at the bases of these jets. Many jets have a clear inverted Y-shape (Figure~\ref{fig.2}). Note that these GST images, as well as the IRIS and SDO/AIA images to be shown later in this paper, have been rotated so that the jets roughly extend upward. In the observations these jets propagate towards the west solar limb. 

Unlike collimated jet-like features, these inverted Y-shaped jets may not be easily explained by magnetohydrodynamic turbulence \citep{Cranmer2015}, warps in two-dimensional sheet-like structures \citep{Judge2011}, or amplified magnetic tension caused by ion-neutral interactions \citep{Martinez-Sykora2017}. These mechanisms were proposed to explain the formation of fast linear jet-like features such as Type-II spicules \citep{DePontieu2007} or TR network jets \citep{Tian2014a}. The double horns at the bases of inverted Y-shape jets in our observation are not necessarily present in these scenarios. Instead, the inverted Y-shape in our GST observation most likely indicates the occurrence of magnetic reconnection in light bridges, thus providing strong support to the following previously proposed scenario \citep[e.g.,][]{Jurcak2006,Lagg2014,Louis2015}: small-scale magnetic bipoles (or a magnetic arcade) in light bridges may reconnect with the unipolar magnetic fields extending from surrounding umbral regions. Unfortunately, an investigation of the magnetic field topology for these small-scale jets is very difficult, as the observed region is rather close to the limb and the nearest in time Hinode data were obtained two days prior to our GST observation, with the light bridge having evolved in the meantime. Another ongoing project of ours is focused on the magnetic field evolution of a light bridge observed with GST around the disk center, which shows possible signatures of frequent occurrence of flux emergence. These data are currently being analyzed and the relevant results will be reported in an upcoming paper.

Since these inverted Y-shaped structures are observed far out in the wings of the H${\alpha}$ line that form in or close to the photosphere, we may take the heights of the horn-like bases of the jets as a measure of the heights where the reconnection occurs. It turns out that for most jets this height is in the range of 250--750 km, indicating that the reconnection takes place in the upper photosphere or lower chromosphere. Considering the height of occurrence, these reconnection events are similar to Ellerman bomb \citep{Ellerman1917}, which are believed to be formed by reconnection taking place around the TMR or in the upper photosphere \citep[e.g.,][]{Ding1998,Watanabe2011,Yang2013,Vissers2013,Nelson2015,Rezaei2015,Rouppe2016,Rutten2016,Nelson2017,Danilovic2017,Libbrecht2017}. It is worth noting that Ellerman bombs are usually observed outside sunspots in the areas of emerging flux and that they are normally believed to be generated by magnetic reconnection at the upper part of U-shaped magnetic structures \citep[e.g.,][]{Isobe2007,Archontis2009,Georgoulis2002,Pariat2009,Cheung2010,Schmieder2014}. However, the frequent occurrence of the inverted Y-shaped jets in our observations indicates intermittent reconnection between small-scale magnetic arcades at the light bridges and unipolar magnetic fields extending from the surrounding umbrae. Despite the totally different magnetic field geometry, both types of reconnection take place in or close to the photosphere and result in efficient energy conversion.

\begin{figure*}
\centering {\includegraphics[width=\textwidth]{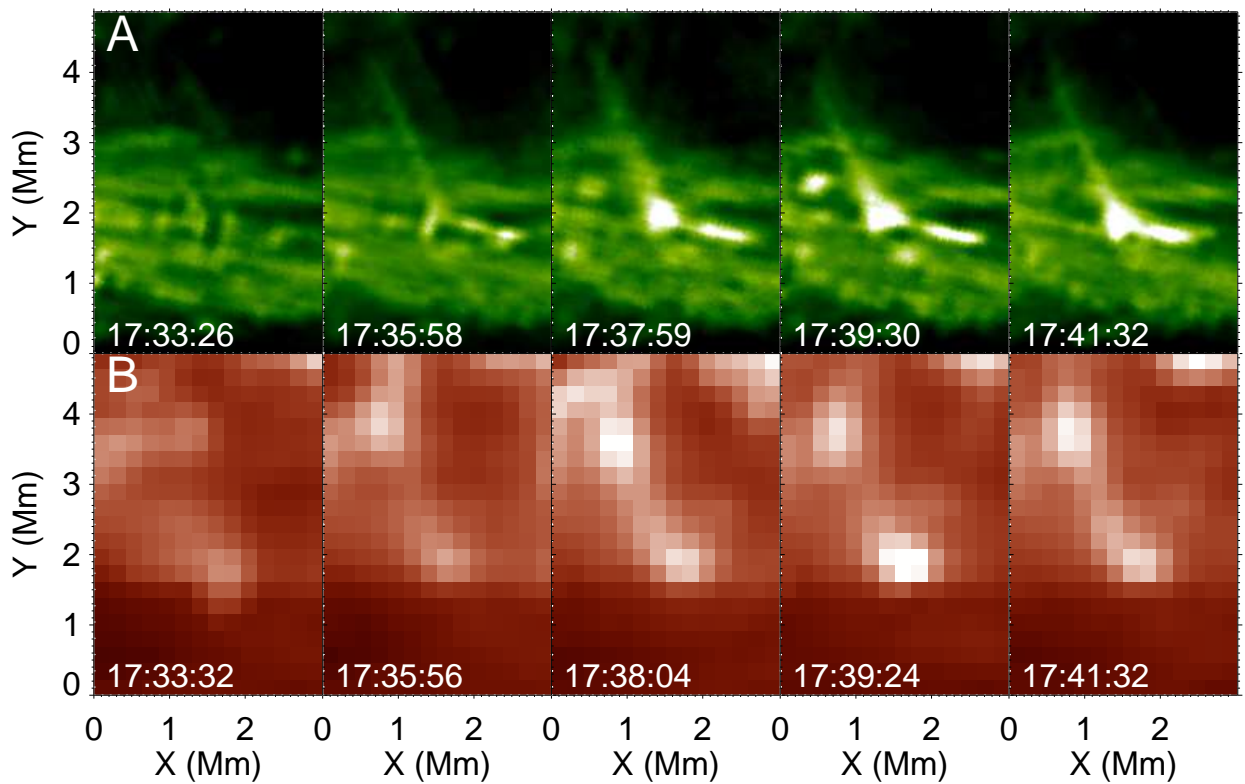}} \caption{ Recurring jets observed around 17:38 UT on 2014 Oct 29.  (A) H${\alpha}$ wing images at --0.8 \AA{} observed by GST. (B) IRIS 1330 \AA{} slit-jaw images. An associated animation (m3.mp4) is available online. } \label{fig.3}
\end{figure*}

Signatures of these jets can also be identified from simultaneous observations of IRIS. IRIS slit-jaw imaging observations with the 1330 \AA{} filter clearly reveal transient brightenings at the footpoints of these jets, indicating heating of the sunspot atmosphere to a few tens of thousands of Kelvin (Figure~\ref{fig.3}). However, the inverted Y-shape is not visible in the IRIS images which have a spatial resolution much lower than that of the GST data. Enhanced 1330 \AA{} emission can also be identified at the tips of the jets, indicating local heating possibly due to shock fronts associated with these fast jets or compression between the jets and the overlying atmosphere.

Many jets tend to recur at roughly the same locations on time scales of minutes, suggesting repeated or intermittent reconnection. For instance, Figure~\ref{fig.3} shows four jets with similar morphology occurring within six minutes. Most jets reach their maximum heights of 4--12 Mm within one or two time steps. Considering the 30-second observing cadence, the apparent speeds of most jets should be 50--400 km~s$^{-1}$ or higher. Such speeds notably exceed the 10--20 km~s$^{-1}$ speeds of the chromospheric anemone jets frequently observed outside sunspots \citep{Shibata2007}. However, sunspots have the strongest magnetic fields on the Sun. If we assume a magnetic field strength of 1500 Gauss and a particle density of 10$^{21}$ m$^{-3}$ near the TMR, then the Alfv\'en speed is estimated to be $\sim$100 km~s$^{-1}$, which is comparable to the observed high speeds and thus reinforcing our conclusion that these jets are produced by magnetic reconnection. 

\cite{Shimizu2009} and \cite{Toriumi2015a} calculated the electric current based on photospheric magnetic field measurements, and found that jets prefer to be initiated from locations where the current is enhanced. This scenario can not be examined in our observation. This is because the observed region is too close to the limb, and thus it is very difficult to tell whether the jets are initiated from one side or both sides of the narrow light bridges. The locations of current enhancement obviously depend on the magnetic environment of the light bridges and the geometry of the surrounding umbral magnetic fields. As we mentioned above, an investigation of the magnetic topology is very difficult in our case. 

The photospheric TiO emission appears to be enhanced at the locations where reconnection jets occur frequently, which may be caused by the reconnection related heating. The H${\alpha}$ core passband is a typical chromospheric passband and persistent surge-like activity is clearly observed above the light bridges. These surge-like events may or may not be related to the inverted Y-shaped jets observed in the H${\alpha}$ wing images, which will be discussed in Section 5.

\section{Reconnection jets and the associated heating: IRIS and AIA observations on 2014 Oct 28}

\begin{figure*}
\centering {\includegraphics[width=\textwidth]{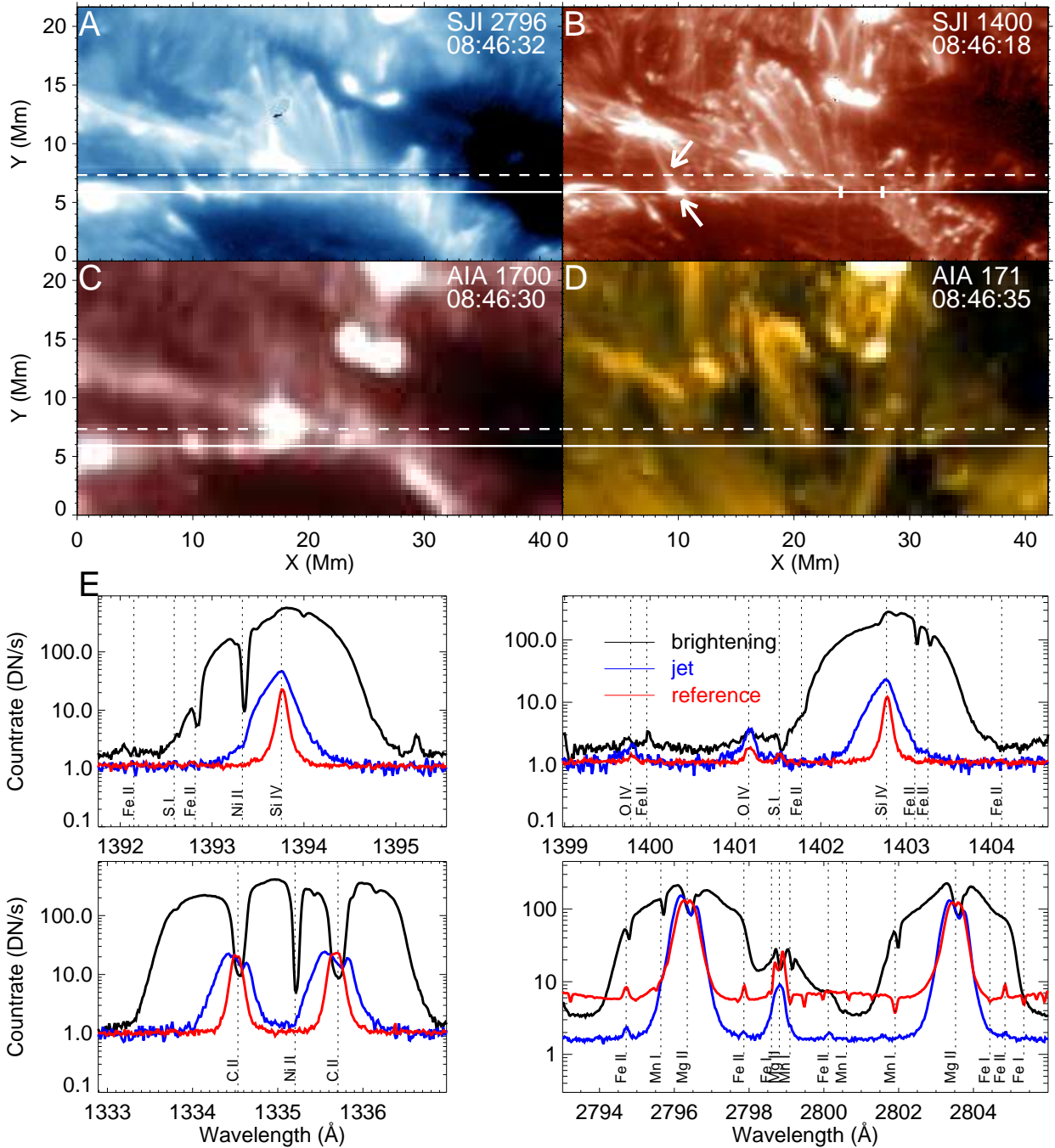}} \caption{IRIS and SDO/AIA observations of a reconnection jet from a light bridge at 08:46 UT on 2014 Oct 28. (A-B) IRIS 2796 \AA{} and 1400 \AA{} slit-jaw images. (C-D) AIA images in the 1700 \AA{} and 171 \AA{} passbands. The IRIS slit positions at 08:46:18 UT and 08:46:32 UT are indicated by the solid and dashed lines, respectively. (E) IRIS line profiles at the footpoint brightening and the jet indicated by the arrows in (B). The reference spectra are the spectra averaged within the section between the two vertical bars in (B). An associated animation (m4.mp4) is available online. } \label{fig.4}
\end{figure*}

The same sunspot was also observed by IRIS on 2014 Oct 28 (Figure~\ref{fig.4}). Numerous collimated jets with footpoint brightenings can be clearly identified from the IRIS 2796 \AA{} and 1400 \AA{} slit-jaw images. Since the 2796 \AA{} and 1400 \AA{} filters sample mainly the emission from plasma with a temperature of $\sim$10,000 K and $\sim$80,000 K respectively, these observations indicate that the jets are multi-thermal and that reconnection may heat local materials to at least $\sim$80,000 K. The AIA 1700 \AA{} filter samples emission mainly from the TMR in the upper photosphere. The compact transient brightenings, which mark the reconnection related heating at the footpoints of the jets, are the most notable features in the 1700 \AA{} images. Weak signatures of several jets can also be identified from these images, possibly suggesting the ejection of the TMR plasma. Most of these jets are associated with dark surges, or are not visible at all in the AIA 171 \AA{} coronal passband, probably due to the obscuration by the foreground coronal structures.

The IRIS slit also crossed a few jets and compact brightenings at the jet footpoints. As exemplified in Figure~\ref{fig.4}, profiles of several ultraviolet emission lines of Si~{\sc{iv}}, C~{\sc{ii}} and Mg~{\sc{ii}} ions sampled at the location of the jet all reveal an enhancement of the blue wing intensities, indicating excess emission from an outward moving gas. The profiles of these spectral lines measured at the compact brightenings are significantly broadened and greatly enhanced at both wings. The wing enhancements extend beyond 200 km~s$^{-1}$ from their rest wavelengths in both directions, suggesting the presence of fast bi-directional flows within a small region \citep{Dere1989,Innes1997,Chae1998,Madjarska2004,Ning2004,Teriaca2004,Huang2014,Huang2017,Peter2014,Gupta2015}. Similar line profiles were previously reported for a repeated transient brightening at a light bridge \citep{Toriumi2015a}.

Another prominent feature in the ultraviolet spectra of the compact brightening is the superposition of several absorption lines on the greatly broadened wings of the Si~{\sc{iv}}, C~{\sc{ii}} and Mg~{\sc{ii}} line profiles. The presence of the Fe~{\sc{ii}} and Ni~{\sc{ii}} absorption lines indicates that the hot reconnection region is located below the cooler chromosphere. In addition, the O~{\sc{iv}} 1401.16 \AA{} and 1399.77 \AA{} forbidden lines are almost absent at the compact brightening, placing the reconnection site in a very dense region \citep{Peter2014}. These characteristics, together with the great enhancement of the extended Mg~{\sc{ii}} wings and the intense brightenings observed in the AIA 1700 \AA{} images, suggest that this reconnection event likely occurs in or just above the upper photosphere \citep{Tian2016,Grubecka2016}. 

\begin{figure*}
\centering {\includegraphics[width=\textwidth]{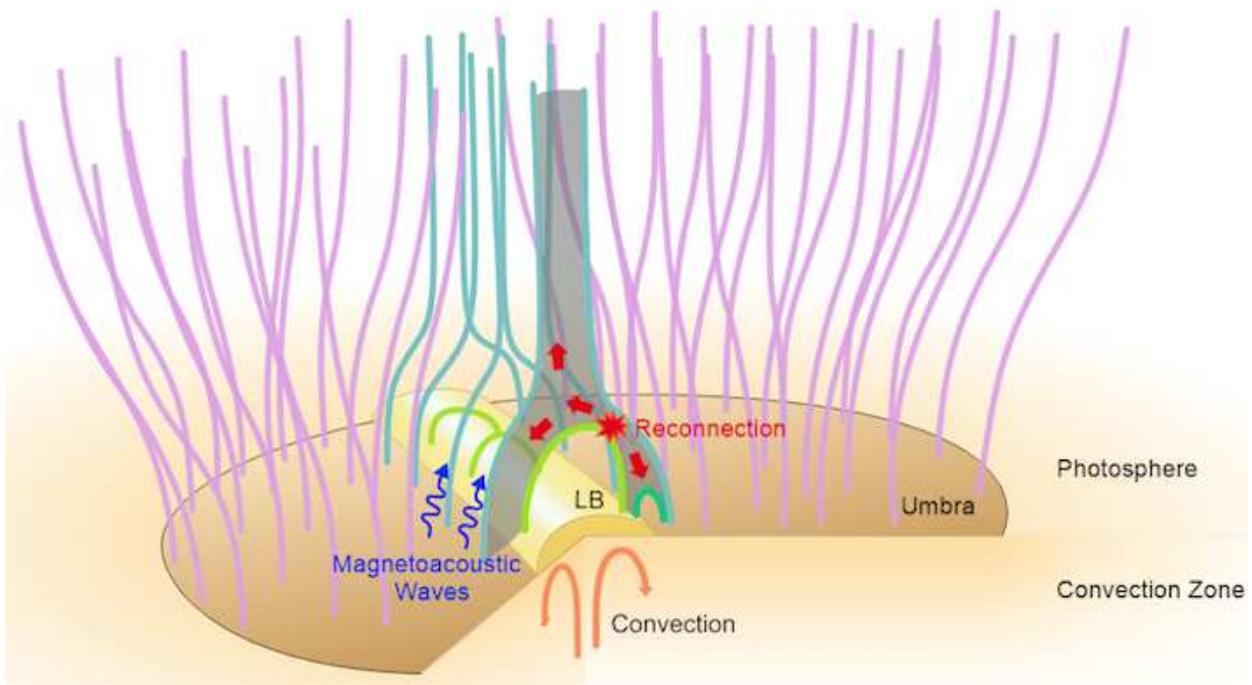}} \caption{ A cartoon illustrating the possible generation process of reconnection jets from light bridges (LB). Low-lying magnetic structures emerge through convective flows from beneath the light bridges. Upward motion of these magnetic structures brings part of them into contact with the overlying strong magnetic field extending from the nearby umbral regions. At one side this field is of opposite polarity to the emerging flux, and a current sheet is formed. The rising magnetic arcade (or bipoles) pushes the field lines at the current sheet together, leading to magnetic reconnection and fast jets with an inverted Y-shape \citep[e.g.,][]{Shibata1994,Shibata2007,Yokoyama1996}. } \label{fig.5}
\end{figure*}

Obviously, the IRIS spectra of the compact brightenings at the jet footpoints are very similar to those of the hot explosions (IRIS bombs or UV bursts) discovered outside sunspots \citep[e.g.,][]{Peter2014,Vissers2015,Kim2015,Tian2016,Chitta2017}. Recent investigations suggest that the hot UV bursts are  formed in the lower atmosphere of emerging ARs, and that at least some of them result from local heating by magnetic reconnection at the upper part of the U shape when a U-loop is dragged down by the accumulated mass \citep{Toriumi2017,Zhao2017,Tian2018}. The jets in our observations are produced through reconnection in totally different magnetic field structures. The inverted Y-shape indicates reconnection between small-scale magnetic bipoles (or a narrow, low-lying arcade) at light bridges and the magnetic canopy rooted in the surrounding umbrae, a scenario previously proposed by some authors \citep[e.g.,][]{Jurcak2006,Louis2015}. These low-lying magnetic structures likely emerge through vigorous convection upflows \citep{Lagg2014,Rimmele1997} from beneath the light bridges \citep{Louis2015,Song2017}. They are restricted to the lower atmosphere by the overlying strong magnetic field of the umbra, which expands to cover the light bridge. Any upward motion of the arcade (or bipoles) brings part of it into contact with the overlying field. At one side this field is of opposite polarity to the emerging arcade, so that a current sheet is produced (see sketch in Figure~\ref{fig.5}). The rising arcade pushes the field lines at the current sheet together, leading to magnetic reconnection. Such a highly dynamic process has not yet been reproduced in any sunspot models, although recently attempts have been made to simulate the small-scale flux emergence process in  light bridges \citep{Toriumi2015b}. Future modeling efforts may need to focus on the interaction between these emerging fluxes and the background sunspot fields. 

The significant heating at the jet footpoints challenges our current understanding of heating by magnetic reconnection around the TMR. Almost all theoretical investigations indicate that the TMR is unlikely to be heated to temperatures exceeding 10,000 K \citep[e.g.,][]{Fang2006,Fang2017,BelloGonzalez2013,Berlicki2014,Hong2014,Hong2017a,Hong2017b,Li2015,Reid2017,Hansteen2017}. Nevertheless, recent magneto-hydrodynamic simulation showed that reconnection near the TMR can indeed heat some of the materials to a temperature of $\sim$80,000 K through shocks provided that the involved magnetic fields are stronger than 500 Gauss \citep{Ni2016}. Sunspots are known to possess such strong magnetic fields, which are usually above 1000 Gauss. Thus, this model may explain the presence of hot materials in our observations. However, this type of investigation needs to be redone, as it lacks a realistic treatment of the radiative cooling and does not consider some important physical processes in the partially ionized sunspot atmosphere such as non-equilibrium ionization. These effects may play important roles in the heating process \citep{Ni2018}.

Systematic wavelength variation during an orbit and along the slit has been removed in the level 2 IRIS data. However, absolute wavelength calibration is still required to derive accurate values of the Doppler shifts of various spectral lines. We average the line profiles within a relatively quiet region of the light bridge. The resultant line profiles are then taken as reference line profiles (the red lines in Figure~\ref{fig.4}) and are used to perform absolute wavelength calibration. For the Si~{\sc{iv}} 1402.77 \AA{} window, the cold chromospheric S~{\sc{i}} 1401.51 \AA{} line in the reference spectrum can be safely assumed to have a zero Doppler shift. For the Si~{\sc{iv}} 1393.76 \AA{} window, cold chromospheric lines do not show strong signals. We thus force the Si~{\sc{iv}} 1393.76 \AA{} and 1402.77 \AA{} lines in the reference spectrum to have the same Doppler shift. The wavelength calibration for the C~{\sc{ii}} window is then achieved by assuming the same Doppler shift for the Ni~{\sc{ii}} 1393.33 \AA{} and 1335.20 \AA{} absorption lines at the compact brightening exemplified in Figure~\ref{fig.4}. For the Mg~{\sc{ii}} window, the photospheric absorption lines Mn~{\sc{i}} 2801.908 \AA{} and Fe~{\sc{i}} 2805.347 \AA{} in the reference spectrum are assumed to be at rest. 

The calibrated spectrum at the compact brightening shows a red shift of a few km~s$^{-1}$ for the Ni~{\sc{ii}}, Fe~{\sc{ii}} and Mn~{\sc{i}} absorption lines. This result appears to be robust, as similar red shifts are obtained when we choose other relatively quiet regions (e.g., a quiet plage region outside the sunspot) to construct the reference spectrum. Using this calibration method, we also find that the Cl~{\sc{i}} 1351.657 \AA{} line appears to have two components, one is nearly stationary and the other is slightly redshifted. The zero Doppler shift of the first component confirms the accuracy of our wavelength calibration. 

The redshifted component is probably caused by the reconnection jet just above the reconnection site. The possible cause of the red shift is the line of sight effect. Let us consider the reconnection configuration shown in Figure~\ref{fig.5}. In the slanted viewpoint the upward branch of the bidirectional jets (oppositely propagating reconnection outflows) in the hot reconnection region may propagate away from the IRIS instrument. In this case the overlying cool material should also show a slow motion away from the instrument, which may explain the redshifted component. At higher layers the upward moving jets will be guided by the umbral field lines that are likely to be more vertical as compared to the reconnection current sheet. Such a geometry may cause the jets to have a velocity component towards the instrument, leading to the blue wing enhancement of the profiles of the Si~{\sc{iv}}, C~{\sc{ii}} and Mg~{\sc{ii}} lines at locations of the jets (Figure~\ref{fig.4}). However, this scenario appears to have difficulty in explaining the redshifted Ni~{\sc{ii}}, Fe~{\sc{ii}} and Mn~{\sc{i}} absorption lines. More investigation should be performed to understand this result in the future.  

%The redshifted component is probably caused by the reconnection jet just above the reconnection site. The most likely cause of these red shifts appears to be the line of sight effect. \textbf{Figure~\ref{fig.5} shows the standard cartoon for inverted Y-shaped jets \citep[e.g.,][]{Shibata1994,Shibata2007,Yokoyama1996}. In the slanted viewpoint (indicated by the purple arrows) the upward branch of the bidirectional jets (oppositely propagating reconnection outflows)} in the hot reconnection region may propagate away from the IRIS instrument. In this case the overlying cool material should also show a slow motion away from the instrument, which may explain the small red shifts of the Ni~{\sc{ii}}, Fe~{\sc{ii}} and Mn~{\sc{i}} absorption lines. At higher layers the upward moving jets will be guided by the umbral field lines that are likely to be more vertical as compared to the reconnection current sheet. Such a geometry may cause the jets to have a velocity component towards the instrument, leading to the blue wing enhancement of the profiles of the Si~{\sc{iv}}, C~{\sc{ii}} and Mg~{\sc{ii}} lines at locations of the jets (Figure~\ref{fig.4}).

\section{Two types of surge-like activity above light bridges}

\begin{figure*}
\centering {\includegraphics[width=\textwidth]{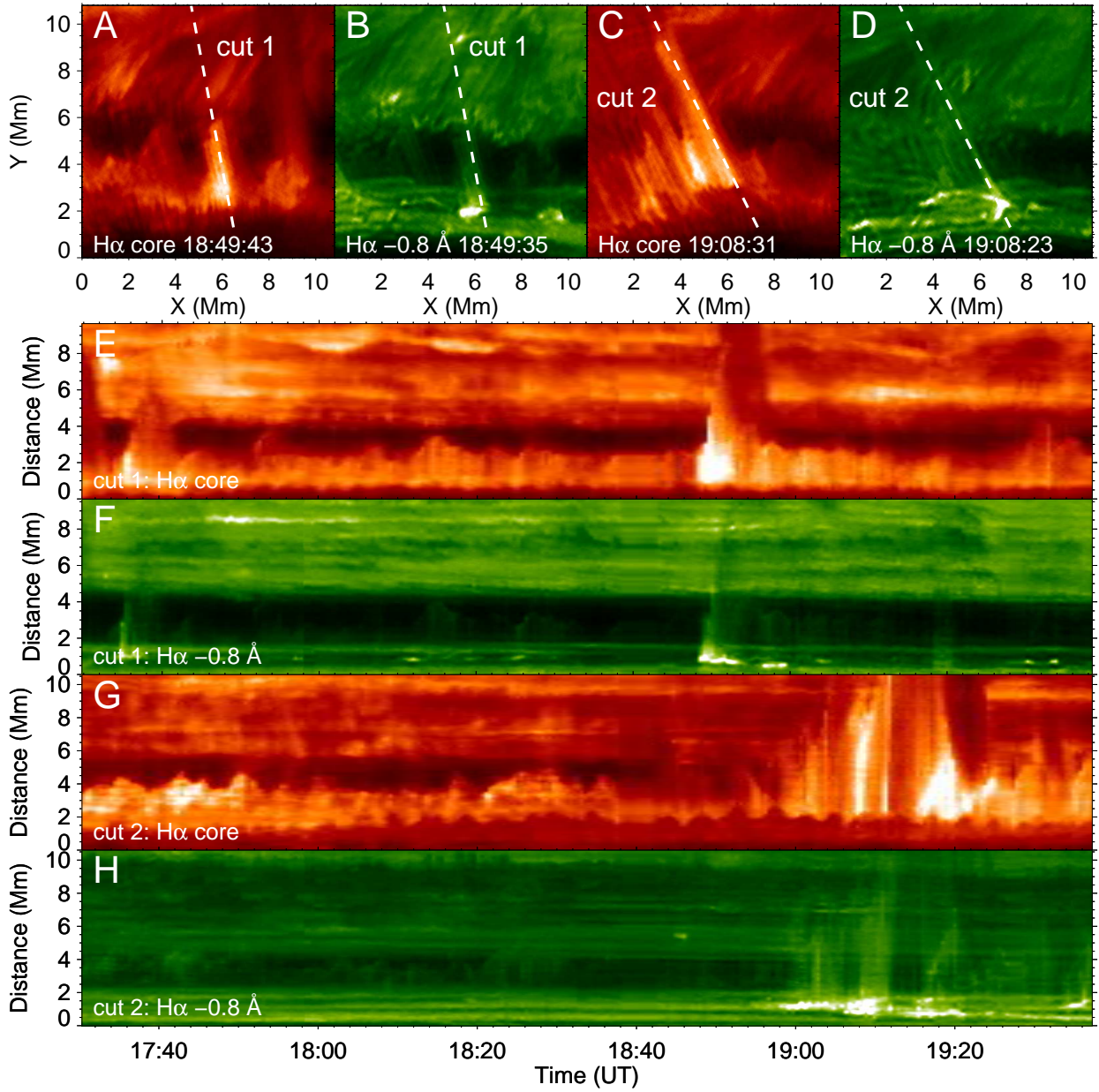}} \caption{ Two types of surge-like activity observed by GST on 2014 Oct 29. (A-D) H${\alpha}$ core and --0.8 \AA{} images taken at 18:49 UT and 19:08 UT. The region shown in (A-B) is different from that in (C-D). (E-H) Space-time diagrams for the two cuts marked in (A-D). } \label{fig.6}
\end{figure*}

\begin{figure*}
\centering {\includegraphics[width=\textwidth]{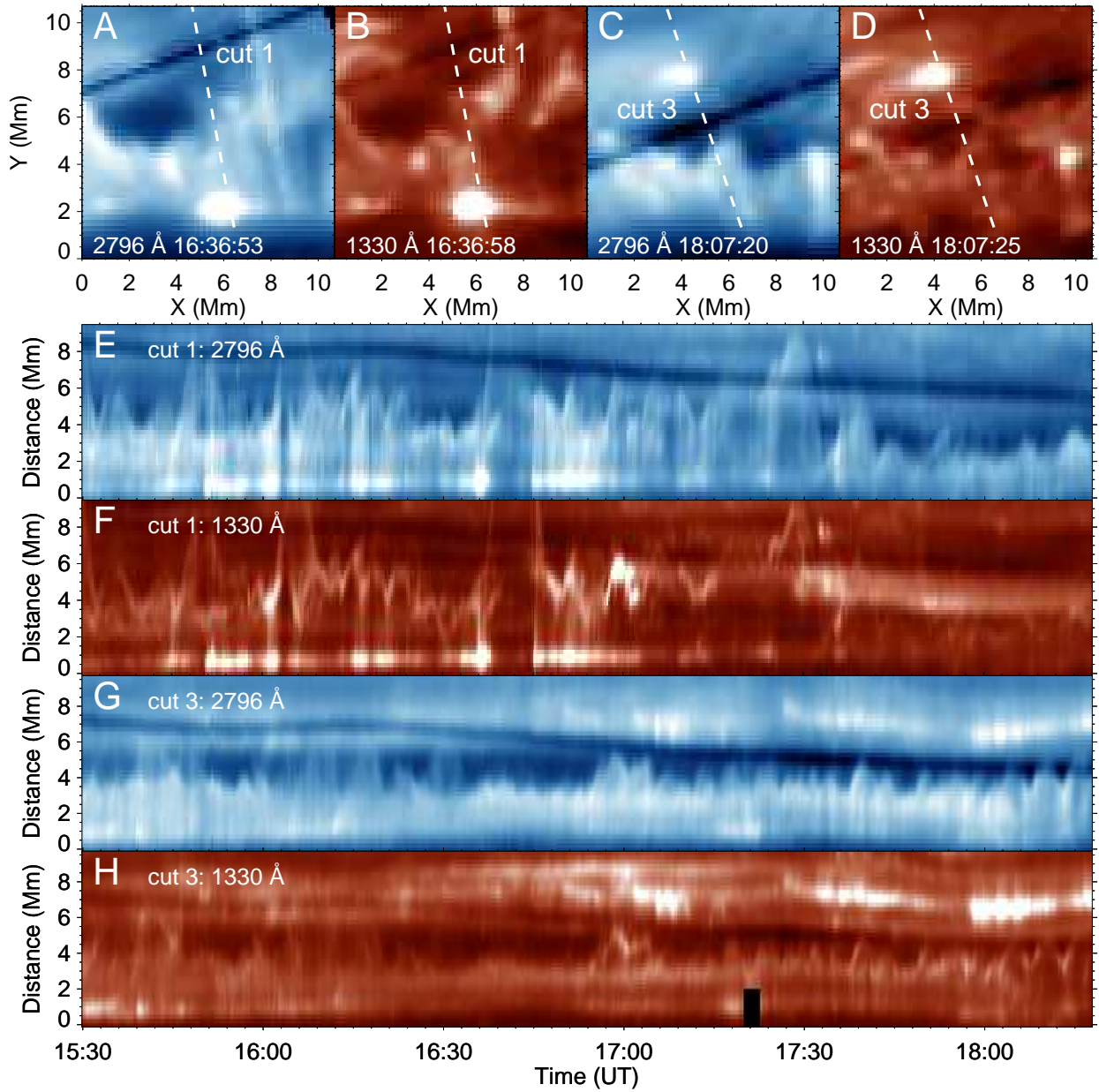}} \caption{Two types of surge-like activity observed by IRIS on 2014 Oct 29. (A-D) 2796 \AA{} and 1330 \AA{} slit-jaw images taken at 16:36 UT and 18:07 UT. The region shown in (A-B) is different from that in (C-D). The regions shown in (A-B) of this figure and (A-B) of Figure~\ref{fig.6} are the same. (E-H) Space-time diagrams for the two cuts marked in (A-D). } \label{fig.7}
\end{figure*}

\begin{figure*}
\centering {\includegraphics[width=\textwidth]{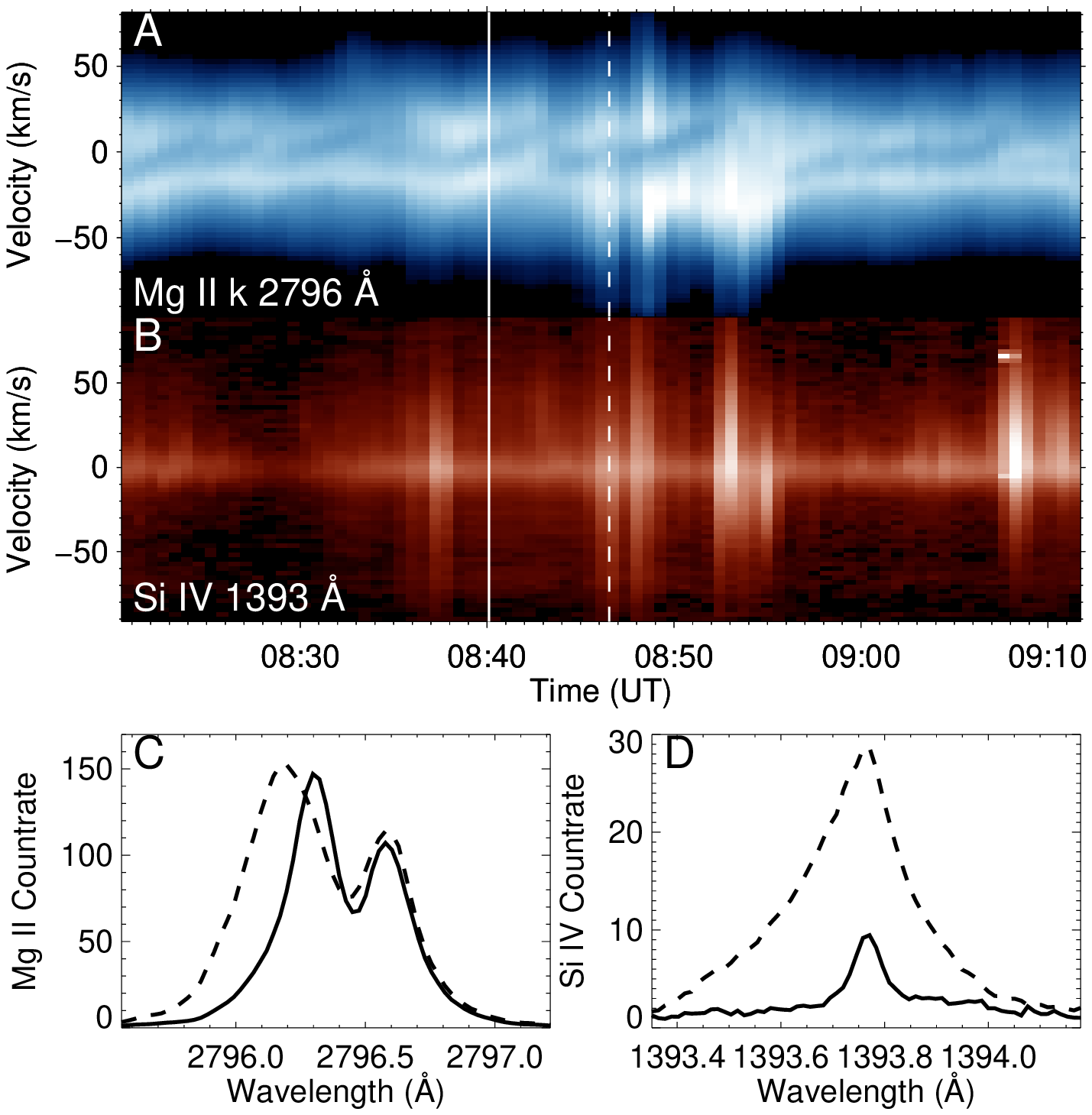}} \caption{ Typical line profiles of Mg~{\sc{ii}} k and Si~{\sc{iv}} 1393.76 \AA{} in the two types of surges observed by IRIS on 2014 Oct 28. (A)-(B) Temporal evolution of the line profiles at the location indicated by the upper arrow shown in Figure~\ref{fig.4}(B). (C)-(D) The solid and dashed curves represent the line profiles taken at the two times indicated respectively by the solid and dashed lines in (A)-(B).} \label{fig.8}
\end{figure*}

As mentioned above, the GST H${\alpha}$ core passband reveals long-lasting recurring surge-like activity, forming an oscillating "wall" above the light bridges. Similar surges are also visible from the IRIS 2796 \AA{} images. Such chromospheric surges were first identified by \cite{Roy1973} and later studied by \cite{Asai2001}. With the high resolution of modern solar telescopes, these surges have been frequently reported and intensively investigated in the past decade \citep[e.g.,][]{Bharti2007,Shimizu2009,Yuan2016,Song2017}. A bright oscillating front has also been identified ahead of the surges from the IRIS 1400 \AA{} and 1330 \AA{} slit-jaw images \citep{Bharti2015,Yang2015}, suggesting heating of the surge front to temperatures of at least 80,000 K that may occur under ionization equilibrium. Recent investigations using the IRIS data showed that this surge-like activity is very common above light bridges \citep{Yang2016,Hou2016a,Hou2016b,Zhang2017}. 

Some surges show high apparent speeds ($\sim$100 km~s$^{-1}$) or coincide with strong currents, leading to the overwhelming interpretation of these surges as reconnection jets  \citep[e.g.,][]{Asai2001,Louis2014,Toriumi2015b,Robustini2016}. However, recent IRIS observations of these surges appear to reveal much lower speeds ($\sim$15 km~s$^{-1}$) and nearly stable oscillating periods of a few minutes, suggesting their cause to be the upward leakage of p-modes from the photosphere \citep{Yang2015,Zhang2017}. 

Our GST observations shed new light on the nature of these chromospheric surges. From Figure~\ref{fig.2} and the associated online animation, these H${\alpha}$ core surges appear to reach higher and become brighter when the reconnection jets are detected at the same locations in the H${\alpha}$ line wings. This suggests the presence of two types of plasma surges above the light bridges. Type-I surges are characterized by constant up-and-down motions with a relatively stable recurrence period as seen in the H${\alpha}$ core data. They appear to occur everywhere along the light bridges and generally reach a height of 0.5-4 Mm. These surges are very similar to those reported by \cite{Zhang2017}. A comprehensive investigation of the dynamics and spectra of these surges led \cite{Zhang2017} to conclude that they are likely driven by shocks that form when p-mode or slow-mode magnetoacoustic waves generated by convective motions propagate upward to the chromosphere. The up-and-down oscillatory motions, parabolic paths and low maximum speeds ($\sim$10--30 km~s$^{-1}$) of these surges appear to be similar to those of the dynamic fibrils observed in plage regions \citep{Hansteen2006,DePontieuHansteen2007} and sunspots \citep{Rouppe2013,Yurchyshyn2014}. A similar scenario has also been proposed by these authors to explain these narrow dynamic fibrils. 

The relatively rare but still frequent Type-II surges are characterized by impulsive ejection of chromospheric material from selected locations at light bridges to heights normally exceeding 4 Mm and often reaching 10 Mm or more. Sometimes many of these long surges occur in succession from the same locations with time gaps of minutes. However, at other times we may not see any Type-II surges over the course of a few hours. They are obviously related to the reconnection jets visible in the H${\alpha}$ wing images, and thus are driven by reconnection between newly emerged magnetic structures in the light bridges and the surrounding umbral magnetic fields. Recently \cite{Robustini2017} observed an emerging loop-like magnetic structure in a penumbra and the subsequent launching of fan-shaped jets. \cite{Magara2010} simulated magnetic reconnection between a horizontal magnetic flux tube and the background field in the penumbra. We think that a similar process may also occur at a light bridge. Figure~\ref{fig.5} illustrates the physical processes involved in these two types of surges. 

The space-time diagrams presented in Figure~\ref{fig.6} clearly reveal these two types of surges. At locations of the two selected cuts, we mainly see short type-I surges during most of the observing period. At cut 1 two type-II surges or reconnection jets occur at around 17:35 UT and 18:49 UT. They reach a height of $\sim$5 Mm and $\sim$8 Mm, respectively. The apparent speeds of these two jets can be estimated from the slopes of the jet trajectories in the space-time diagrams, which turn out to be $\sim$50 km~s$^{-1}$ in both cases. At cut 2 several type-II surges occur intermittently after 19:00 UT. The trajectories for most of these surges appear to be vertical in the space-time diagrams, meaning that these jets reach their maximum heights very quickly, usually within one or two time steps. As mentioned in the main text, their apparent speeds are estimated to be 50--400 km~s$^{-1}$ or higher.

These two types of surges are also revealed in the space-time diagrams using the IRIS slit-jaw images, as shown in Figure~\ref{fig.7}. Frequent occurrence of compact brightenings can be identified at the crossing between cut 1 and the light bridges, marking intense reconnection-driven activity. Correspondingly, we see many high-reaching Type-II surges in the space-time diagrams. After 17:36 UT, there are no significant brightenings at the footpoints of the surges any more, suggesting the end of reconnection activity. As a result, short Type-I surges re-appear. Note that there is a time overlap between the GST and IRIS observations from 17:30 UT to 18:18 UT. Similar surges are revealed from the space-time diagrams of H${\alpha}$ core (Figure~\ref{fig.6}) and 2796 \AA{} (Figure~\ref{fig.7}) images for cut 1 during this period. In contrast, at cut 3 we only see the short Type-I surges with a recurrence period of $\sim$4 minutes. And there is almost no significant brightening at the light bridge during the entire observing period, indicating the absence of reconnection at this location. 

We have also compared the Mg~{\sc{ii}} k and Si~{\sc{iv}} 1393.76 \AA{} line profiles within the two types of surges. Figure~\ref{fig.8} presents the temporal evolution of the line profiles observed at one location above a light bridge on 2014 Oct 28. Several Type-II surges are clearly present during the observation period. These surges are characterized by obviously enhanced and broadened Si~{\sc{iv}} line profiles, indicating that these surges reach a temperature of the order of $\sim$80,000 K. For most of these surges, the Mg~{\sc{ii}} k line also shows an enhancement in the intensity, especially in the blue wing intensity. The Si~{\sc{iv}} line in the ever-present Type-I surges is often a few times weaker and much narrower than in Type-II surges. While the Mg~{\sc{ii}} k line has strong emission in these Type-I surges. \cite{Zhang2017} concluded that the persistent Type-I surges mainly consist of chromospheric plasmas and that only the bright fronts ahead of the surges are heated to typical TR temperatures. This scenario appears to be consistent with our spectroscopic observations. In the IRIS slit-jaw imaging observations of Type-I surges, we also often see obvious emission in the 2796 \AA{} filter. On the contrary, the 1330 \AA{} and 1440 \AA{} filters which sample mainly TR emissions usually reveal reduced emission in the Type-I surges and enhanced emission ahead of these surges. This observational fact also suggests that Type-I surges have a temperature of $\sim$10,000 K, thus supporting the conclusion of \cite{Zhang2017}. Figure~\ref{fig.8}(A) also reveals hints of sawtooth pattern in the temporal evolution of the Mg~{\sc{ii}} k line profile. The dark line core appears to show episodes of the following behavior: a rapid excursion to the blue followed by a gradual shift to the red at a constant rate. A similar pattern has been previously reported for shock dominant phenomena such as dynamic fibrils \citep[e.g.,][]{Langangen2008} and sunspot oscillations \citep[e.g.,][]{Rouppe2003,Tian2014b,Yurchyshyn2015}. The constant deceleration/acceleration in each episode is consistent with the parabolic path that is usually observed in Type-I surges, as mentioned above. It should be noted that the sawtooth pattern in our observation is not as clear as in these previous investigations, possibly because our observed region is too close to the limb. Nevertheless, Figure~\ref{fig.8}(A) still provides support for the conclusion that the ever-present Type-I surges are likely upwardly propagating shock waves.

The driving mechanisms of these two types of surges appear to be similar to those of the chromospheric spicules observed outside sunspots. Based on off-limb chromospheric observations, \cite{DePontieu2007} classified spicules into two types. Type-I spicules usually show up-and-down motions with a period of 3--7 minutes and they are driven by shock waves that form when magnetoacoustic waves leak into the chromosphere. The much more dynamic and short-lived Type-II spicules often show only upward motions with speeds of 50--150 km~s$^{-1}$, and they have been proposed to be driven by magnetic reconnection by \cite{DePontieu2007}. It appears clear that the Type-I spicules and our Type-I light bridge surges are driven by a similar physical process. However, a recent study by \cite{Martinez-Sykora2017} suggests that the fast Type-II spicules likely result from the violently released magnetic tension, which is amplified and transported upward through ion-neutral interactions or ambipolar diffusion. While the inverted Y-shape structures in our H${\alpha}$ wing images and the accompanied footpoint brightenings in the IRIS TR images suggest reconnection to be the most likely nature of our Type-II light bridge surges. 

Contrary to many previous studies, our GST and IRIS observations clearly reveal that the persistent chromospheric surge-like activity above light bridges is not likely to be related to magnetic reconnection. However, occasionally occurring reconnection jets are often superimposed on these gentle and persistent surges, forming a sporadic but more violent component of the surges. By analyzing the AIA and IRIS slit-jaw images of two light bridges, \cite{Hou2017} recently proposed that p-mode waves and magnetic reconnection may simultaneously shape the light bridge dynamics, similar to the scenario described above. They identified intermittent plasma ejections superimposed on an oscillating light wall in the 1400 \AA{} images. The persistent oscillating wall usually rises with a projected speed of $\sim$10 km~s$^{-1}$ and reaches a height of a few Mm; it obviously belongs to the Type-I surges we described above. They also found that the wall top is brighter in the ascending phase than in the falling phase, which also appears to be the case in many of our Type-I surges. They conjectured that the oscillatory motions of the wall are driven by leaked p-mode waves. Besides the global p-mode, we think that slow-mode magnetoacoustic waves generated by convective motions in the photosphere may also play a role in launching these Type-I surges. The intermittent plasma ejections in the observations presented by \cite{Hou2017} reach much larger heights and are often accompanied by footpoint brightenings in the 1600 \AA{} and 1400 \AA{} images, similar to our Type-II surges. However, the speeds of these ejections are found to be $\sim$30 km~s$^{-1}$, which appears to be much smaller than, or at least at the lower end of the apparent speeds of our Type-II surges. Also, these ejections often reach tens of Mm, which is larger than the traveling distances of most Type-II surges in our observations. These differences may be understood in terms of different magnetic free energy and/or line of sight projection effects in different light bridges. In addition, due to the relatively low spatial resolution of the AIA and IRIS data, \cite{Hou2017} could not tell whether inverted Y-shaped structures are present in their light bridges. Since the light bridges analyzed by \cite{Hou2017} were not crossed by the IRIS slit, a comparison between the IRIS spectra at the footpoints of these plasma ejections and our Type-II surges is not possible, either. 

It is worth mentioning that magnetic reconnection occurring near the photosphere could also generate slow-mode waves. These waves can then propagate upward and develop into shock waves \citep{Takasao2013,Yang2014,Song2017}. A similar idea was proposed by \cite{Shibata1982} more than 30 years ago, though in their simulation the source of the waves is an arbitrary sudden pressure enhancement in the low atmosphere. In our observations some high-reaching surges that are proceeded by footpoint brightenings also show a parabolic trajectory in the space-time diagrams. Such Type-II surges may be caused by the passage of shock waves that are generated by reconnection. So the Type-II surges could be the reconnection outflows accelerated by the Lorentz force, or elevated chromospheric and TR plasmas when the reconnection driven shocks pass through.

\section{Summary}

With high-resolution observations in both ultraviolet and visible light, we report detection of frequently occurring magnetic reconnection and the resultant significant heating in the lower atmosphere of sunspots. With IRIS observations, we find many narrow and collimated jets initiating from transient compact brightenings at the sunspot light bridges. Our GST observations reveal an inverted Y-shape for many of these jets, providing strong evidence of magnetic reconnection in sunspots. Our observations provide strong support to the following scenario: relentless convective up-flows below sunspot light bridges continuously transport magnetic fields up to the tenuous atmosphere, where they are destined to reconnect with strong, nearly vertical background fields extending from the umbrae clamping the light bridge. The H${\alpha}$ images and ultraviolet spectra suggest that reconnection in or close to the photosphere efficiently heats the weakly ionized cool materials to at least $\sim$80,000 K. Such highly dynamic jets and the efficient heating have never been seen in any sunspot models. By virtue of the unique magnetic geometry and dark background of sunspots, these jets also open a new window to the study of reconnection in the strong magnetic field environments that are common in astrophysical plasmas. 

Our observations also shed new light on the nature of the previously discovered surge-like activity above some light bridges. These chromospheric surges are clearly seen from our data. Our GST and IRIS observations show that there are actually two types of surges above light bridges: the Type-I persistent surges show up-and-down oscillating motions with a period of a few minutes and are present essentially everywhere above the light bridge. They usually reach heights of 0.5-4 Mm. While the intermittent Type-II surges appear at only selected locations of the light bridges and are clearly related to the reconnection jets mentioned above. They are accompanied by impulsive brightenings at the footpoints and often reach a height beyond 4 Mm. These reconnection jets appear to form a sporadic but more violent component of the surge-like activity above light bridges.

\begin{acknowledgements}
BBSO operation is supported by NJIT, US NSF AGS-1250818 and NASA NNX13AG14G grants. The GST operation is partly supported by the Korea Astronomy and Space Science Institute (KASI), Seoul National University, and the Strategic Priority Research Program of CAS with Grant No. XDB09000000. Authors thank BBSO staff for their help during the observations. IRIS is a NASA Small Explorer mission developed and operated by LMSAL with mission operations executed at NASA Ames Research center and major contributions to downlink communications funded by ESA and the Norwegian Space Center. The authors are supported by NSFC grants 11790304, 41574166, 11573064, 11729301, 11573012 and 11790300, the Recruitment Program of Global Experts of China, the Max Planck Partner Group program, NASA grant NNX15AF48G, the European Research Council (ERC) under the European Union's Horizon 2020 research and innovation programme (grant agreement No. 695075), the BK21 plus program through the National Research Foundation (NRF) funded by the Ministry of Education of Korea, the ISSI Bern international team "Solar UV bursts - a new insight to magnetic reconnection", the AFOSR FA9550-15-1-0322 and NSF AST-1614457 grants and KASI. 
\end{acknowledgements}

\end{document}